\documentclass[11pt,a4paper]{article}
\usepackage{bm}
\usepackage{latexsym}
\usepackage{dcolumn}
\usepackage{amsmath,amsfonts,amssymb}
\usepackage{graphicx,epsfig}
\usepackage{psfrag}
\usepackage{color}
\usepackage{nccmath}
\usepackage{moresize}
\usepackage{enumerate}
\usepackage{subcaption}
\usepackage{epstopdf} 
\usepackage{hyperref}
\usepackage{cite}
\usepackage{stackengine}
\usepackage{scalerel}

\newcommand{\be}{\begin{equation}}
\newcommand{\ee}{\end{equation}}
\newcommand{\bea}{\begin{eqnarray}}
\newcommand{\eea}{\end{eqnarray}}
\newcommand{\bse}{\begin{subequations}}
\newcommand{\ese}{\end{subequations}}
\newcommand{\bce}{\begin{center}}
\newcommand{\ece}{\end{center}}
\newcommand{\bfg}{\begin{figure}}
\newcommand{\efg}{\end{figure}}
\newcommand{\bit}{\begin{itemize}}
\newcommand{\eit}{\end{itemize}}
\newcommand{\bed}{\begin{description}}
\newcommand{\eed}{\end{description}}
\newcommand{\ben}{\begin{enumerate}}
\newcommand{\een}{\end{enumerate}}
\newcommand{\nn}{\nonumber}

\newcommand{\fr}{\frac}
\newcommand{\sq}{\sqrt}
\newcommand{\no}{\noindent}

\def\le {\left}
\def\ri {\right}
\def\d  {\delta}

\def\l  {\lambda}

\def\m  {\mu}

\newcommand{\bdm}{\begin{displaymath}}
\newcommand{\edm}{\end{displaymath}}

\begin{document}

\markboth{Saurya Das, Mitja Fridman and Sourav Sur}
{A novel violation of the equivalence principle}

\title{\LARGE{A novel violation of the equivalence principle}}

\author{Saurya Das\footnote{Email: saurya.das@uleth.ca} ~and  
Mitja Fridman\footnote{Email: fridmanm@uleth.ca}\\ 
{\normalsize \em Theoretical Physics Group and Quantum Alberta, Department of Physics and Astronomy,}\\
{\normalsize \em University of Lethbridge, 4401 University Drive, Lethbridge, Alberta T1K 3M4, Canada}\\ \\
Sourav Sur\footnote{Corresponding Author. Email: sourav@physics.du.ac.in} \\ 
{\normalsize \em Department of Physics and Astrophysics, University of Delhi, Delhi - 110007, 
India}}

\date{}
\maketitle


\begin{abstract}
It is generally assumed that any discrepancy between an object's inertial and gravitational masses, leading to a violation of the equivalence principle, arises from the nature of its internal constituents and their interactions. We show here that the difference can instead be a function of the distance of the object from a gravitating body, and suggest ways of testing this, illustrating side-by-side a covariant framework for the same. 
\end{abstract}

\vspace{10pt}
\no



\section{Introduction} \label{sec:intro}

The Einstein's Equivalence Principle (EEP), which is a cornerstone of General Relativity (GR), states that the gravitational and inertial masses of any object are equal, ensuring all objects in free fall to have the same acceleration irrespective of their masses. 
In fact, the EEP provides the foundation for the local flatness criterion, which leads to the derivation of the geodesic equation and Einstein's field equations 
\cite{einstein-1922,weinberg-1972}.
Nevertheless, there is a reasonable expectation that the EEP may be violated, however mildly, under appropriate conditions 
\cite{will-LRR-2014,will-2018}. 

In this article, we demonstrate that the EEP violation can happen in a novel manner, depending on the distance of test objects from a gravitating body, unless the gravitational forces become negligible, such as that at asymptotically large distances. While the current experiments on EEP cannot rule out this possibility, we suggest concrete ways to test it, referring to its embedding in a covariant theory as well. In fact, our work emphasizes the very need to test the law of gravity so as to have potential violations of the EEP at short distances, beyond the sub-millimeter scale in particular, up to which Newtonian gravity has been experimentally confirmed 
\cite{HKHAGSS-submm,HWWS-submm,WHPA-submm,LACFH-submm}.

\section{Modified gravitational force law and the violation of EEP} \label{sec:EEPviol}

Consider, for simplicity, a system of three point particles --- a central, `gravitating' one, $O_0$, with large gravitational and inertial masses $m_{G0}$ and $m_{I0}$ respectively, and two sufficiently lighter (`test') ones, $O_1$ and $O_2$, with gravitational and inertial masses $\{m_{G1}, m_{I1}\}$ and $\{m_{G2}, m_{I2}\}$ respectively.

Tests for the EEP are designed to put bounds on the so-called `E\"otv\"os ratio' which is defined for the two objects $O_1$ and $O_2$ as 
\cite{will-LRR-2014,will-2018}
\be \label{eta1}
\eta =\, 2 \fr{\le|a_1 - a_2\ri|}{\le|a_1 + a_2\ri|} \,,
\ee
where 
\be \label{accl1}
a_i =\, \fr{G\, m_{G0} m_{Gi}}{m_{Ii}} \,\Phi(r) \,; \quad (i = 1, 2) \,, 
\ee 
denotes their accelerations when they are assumed to be located at the same distance from $O_0$, with $G$ being the Newton's constant, and the function $\Phi(r)$ proportional to the gravitational field ($= 1/r^2$ in Newtonian theory). 

If the objects $O_1$ and $O_2$ are made of different materials, say brass and glass, Eq.\,(\ref{eta1}) implies 
\be \label{eta2}
\eta =\, 2 \le| \fr{m_{G1}}{m_{I1}} - \fr{m_{G2}}{m_{I2}} \ri|\bigg/\le|\fr{m_{G1}}{m_{I1}} + \fr{m_{G2}}{m_{I2}}\ri| \,.
\ee 
Consequently, even a small difference in the ratios of their gravitational to inertial masses would manifest as a small but nonzero $\eta$, entirely attributable to their internal composition. Note however that $\eta$ does not depend on the function $\Phi(r)$, and from the usual (Newtonian or GR) perspective, the gravitational masses being constants, $\eta$ is $r$-independent. 

Nevertheless, since the gravitational mass is effective only when gravitational forces are at play, we explore the possibility of having the gravitational force between $O_0$ and any of the test objects $O_i$ ($i=1,2$) as
\be \label{grav1}
F_i (r) =\, -\, G\, m_{G0}(r)\, m_{Gi} (r)\, \Phi(r) \,, 
\ee
where we propose that the gravitational masses depend on the radial coordinate $r$.  
In particular, our assertion of the functional form of the gravitational mass $m_G$ of any given object is based on the criterion that the EEP remains valid at very large distances, and also possibly at very short distances, while allowing for potential violations in the intermediate range. In other words,
%
\be 
m_G \to m_I \,, ~~ \text{as}~~ r \to \infty \,, \label{mg1}
\ee
where $m_I$ denotes the corresponding inertial mass, and possibly,
\be 
m_G \to m \,, ~~ \text{as}~~ r \to 0 \,, \label{mg2}
\ee
where $m$ is a constant which is {\em fixed} for all objects.
%
%
%
%
A functional form such as the following suffices, for all the three objects $O_0, O_1$ and $O_2$:
%
\be \label{mg3}
m_{Gi} =\, m +\, \fr{(m_{Ii} - m)\,r/\l}{1 + r/\l} \,; \quad (i = 0, 1, 2) \,,
\ee
%
where $\l$ is a characteristic length scale that determines whether $r$ is to be considered large or small, dictating whether Eq.\,(\ref{mg1}) or Eq.\,(\ref{mg2}) applies. 
Note that Eq.\,(\ref{mg3}) implies that two objects with the same gravitational mass will also have the same inertial mass. However, for each individual object, the difference between the gravitational and inertial masses depends on $r$.

It follows from Eqs.\,(\ref{grav1}) and (\ref{mg3}), that for the test objects $O_i$ ($i=1,2$)
\be \label{grav2}
F_i(r) =\, -\, G \le[m + \fr{m_{I0} - m}{1 + \l/r} \ri] \le[m + \fr{m_{Ii} - m}{1 + \l/r} \ri] \Phi(r) \,, 
\ee
and the dependence on $r$ cannot be absorbed by a redefinition of $G$ or $\Phi(r)$. The EEP is clearly violated, in general, as the corresponding acceleration, given by 
\bea 
&& a_i (r) = \fr{|F_i (r)|}{m_{Ii}} \nn\\
&& = G \le[m + \fr{m_{I0} - m}{1 + \l/r}\ri] \le[\fr m {m_{Ii}} + \fr{1 - m/m_{Ii}}{1 + \l/r} \ri] \Phi(r) \\
&& \xrightarrow[r \to \infty]{} G m_{I0} \le[1 + \le(\fr m {m_{I0}} + \fr m {m_{Ii}} - 2\ri) \fr \l r + \dots \ri] \Phi(r) \,, \label{accel1} \\
&& \xrightarrow[r \to 0]{} G m \le[\fr m {m_{Ii}} + \le(1 - \fr{2m}{m_{Ii}} + \fr{m_{I0}}{m_{Ii}}\!\ri) \fr r \l + \dots \ri] \Phi(r) \,, \label{accel2} 
\eea
depends on the respective inertial mass $m_{Ii}$, at any finite value of $r$ (in units of $\l$). In the asymptotic limit $r \to \infty$ though, the EEP holds, as per the criterion.
%

Several points merit attention here:
\bed 
\item Firstly, even in the large $r$ limit, corrections to the standard $m_{Ii}$-independent acceleration arise, due to the sub-leading terms in Eq.\,(\ref{accel1}), which can in principle be tested using gravitational wave detectors 
\cite{SY-GW}.
\item Secondly, in the small $r$ limit, the leading term itself depends on $m_{Ii}$.
\item Thirdly, $m$ acts as a {\em universal}, but yet undetermined, mass scale, introducing an additional length scale $\bar\l \sim 1/m$, which we shall discuss further, alongwith the other length scales involved in our proposal, in the concluding section.
%
\item Finally, the E\"otv\"os ratio $\eta$ for the two test particles $O_1$ and $O_2$ attracted by the gravitating body $O_0$, although independent of $\Phi(r)$, depends explicitly on the distance $r$, in contrast to the $r$-independent relation (\ref{eta2}). This is in fact our key new result, given by  
\eed
\bea 
&&\eta \, (m_{I1},m_{I2},r) = \fr{2 m \le|m_{I1} - m_{I2}\ri|}{m \le(m_{I1} + m_{I2}\ri) + 2 m_{I1} m_{I2}\, r/\l} \label{eta3} \\
&&\xrightarrow[r \to \infty]{} \fr{m \le|m_{I1} - m_{I2}\ri|}{m_{I1} m_{I2}} \fr \l r + {\cal O} \le(\fr 1 {r^2}\ri), \\
&&\xrightarrow[r \to 0]{} \fr{2 \le|m_{I1} - m_{I2}\ri|}{m_{I1} + m_{I2}} \le[1 - \fr{2 m_{I1} m_{I2}}{m \le(m_{I1} + m_{I2}\ri)} \fr r \l\ri] + {\cal O} (r^2). 
\eea 
%

The question that remains now is: how to test this? The simplest way would be to take two test bodies, made of the {\it same} material but having different inertial masses $m_{I1}$ and $m_{I2}$, subject them to gravitational forces from a much heavier 
body, measure their accelerations to determine a non-zero $\eta$ and put bounds on it. 
%
\begin{figure}[!htb]
\captionsetup[subfigure]{
  labelformat=empty, 
  labelsep=period 
}
\centering
\begin{subfigure}{0.48\linewidth} \centering
\includegraphics[scale=0.825]{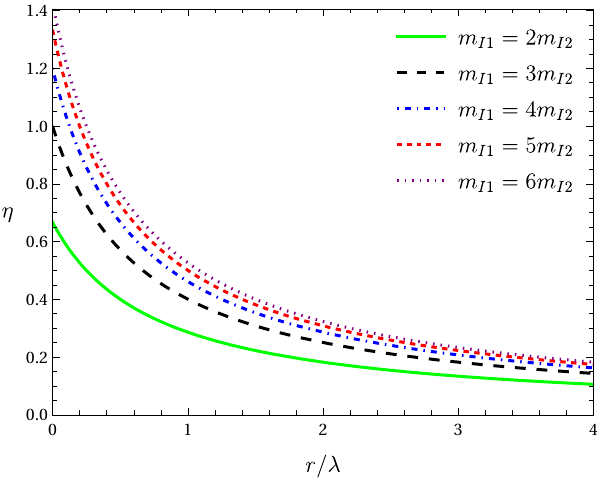}
\caption{\footnotesize {\bf Fig.\,1}: Dependence of $\eta$ on $r/\l$, setting $m = 1$ (in suitable units) and $m_{I1}$ as certain integral multiples of $m_{I2}$.} 
\end{subfigure} ~~
\begin{subfigure}{0.48\linewidth} \centering
\includegraphics[scale=0.85]{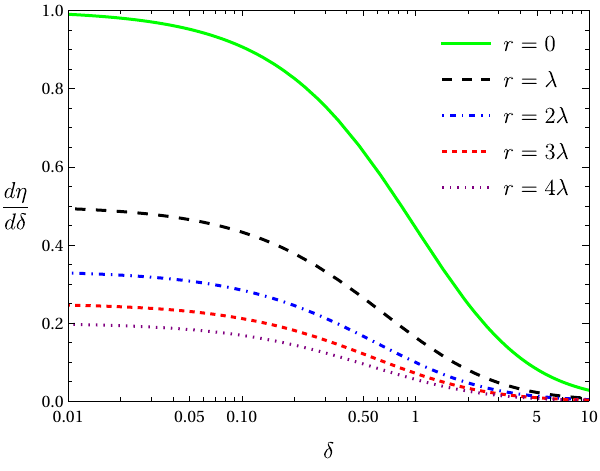}
\caption{\footnotesize {\bf Fig.\,2}: Variation of $d\eta/d\d$ (for $m = 1$) with $\d$ (in a log scale), at certain values of $r$ (in units of $\l$) including $r = 0$.} 
\end{subfigure} 
\label{}
\end{figure}
%
The next step would be to repeat the experiment for different values of $r$ (in units of $\l$). If the corresponding plot of $\eta$ versus $r/\l$ (for a suitable choice of units, say, with $m = 1$) resembles one of those in Fig.\,1, depending on the ratio $m_{I1}/m_{I2}$, it would be an almost irrefutable piece of evidence supporting our proposal. 

Apart from showing that $\eta$ becomes very small for large $r/\l$ and approaches a significantly higher non-zero value as $r \to 0$, for any given ratio $m_{I1}/m_{I2}$, a close insight to Fig.\,1 reveals an interesting feature. That is, with the increase of $m_{I1}/m_{I2}$, the corresponding plots have their adjacent intervening spaces diminishing progressively, at any given value of $r/\l$. This can be utilized in constraining $\eta$ from its variation with the fractional change of the inertial mass of one body relative to that of the other, viz. 
\be \label{delta}
\d =\, \fr{m_{I1} - m_{I2}}{m_{I2}} \,,
\ee 
or more appropriately, that of the slope $d\eta/d\d$ with $\d$ at a particular $r/\l$.  
Fig.\,2 shows the $d\eta/d\d$ versus $\d$ plots, for $m = 1$, at discrete values of $r/\l \in [0, 4]$, and for a wide range of $\d \in [0.01, 10]$. As is seen in all these plots, $d\eta/d\d$ decreases with increasing $\d$, initially slowly, then rapidly, and slowly again, before becoming negligible for a large enough $\d$, whence one of the bodies (that with inertial mass $m_{I2}$ here) merely attains the status of a test body in comparison to the other (the one with inertial mass $m_{I1}$). Note also that at particular value of $\d$, the value of $d\eta/d\d$ gets reduced more and more as $r$ increases (in units of $\l$), just as what happens to the value of $\eta$ with increasing $r/\l$ at a given value of the ratio $m_{I1}/m_{I2}$, i.e. of $\d$ (see the Fig.\,1 above).  

Concrete examples of experiments, for testing our proposal, can be the torsion balance ones and those done in satellites 
\cite{will-LRR-2014,will-2018,smith-etal-eptest,rosi-etal-eptest,TMR-sat-eptest}.
As noted in such examples, one has to replace the two objects of different materials but of the same gravitational mass with two objects made of the {\it same} material but of different gravitational masses. This would hold for other experiments to test the EEP violation as well. 

\section{Modified gravitational force law in a covariant formulation} \label{sec:cov}

In order to embed the above proposal in a covariant theory, it is necessary to specify the form of $\Phi(r)$. For concreteness, we adopt the form we proposed in a recent work
\cite{SDMFSS-PRD-2025},
viz.,
\bea
V(r) &=& - \fr{G m_{G0}}{\ell} \, \fr{(r/\ell)^n}{1 + (r/\ell)^{n+1}} \,, \quad n > 0 \,,
\label{pot}\\
\Phi(r) &=& - \fr 1 {Gm_{G0}}\, \fr{dV}{dr} \nn\\
&=& - \fr{G m_{G0}}{\ell} \le[\fr{(n+1) (r/\ell)^{2 n}}{\le\{(r/\ell)^{n+1}+1\ri\}^2} - \fr{n (r/\ell)^{n-1}}{(r/\ell)^{n+1}+1}\ri] \,,  \label{Phi}
\eea
where $\ell$ is another characteristic length scale (see the discussion below in the concluding section).

For the case above, Newtonian gravity and standard GR get recovered at large distances, whereas at short distances the singularity problem (as $r \to 0$) gets resolved. The corresponding metric component 
\be
g_{00} \simeq\, - \le(1 + 2V\ri) \,,
\ee
can be identified, for small $r$, with that in a covariant modified gravitational theory, specifically, the $f(R)$ theory of gravity of a particular sort, where $R$ denotes the curvature scalar. This is described by an action
\be 
S = \fr 1 {8\pi G} \int d^4 x \sq{-g}\, f(R) \,,
\ee 
with 
\be 
f(R) =\, R +\, \bar c \, R^{2(n-1)/(n-2)} \,, 
%
\ee
where 
\be 
\bar c = \fr{(n-2) \le(G m_{G0}/\ell^{n+1}\ri)^{2/(n-2)}}{2 \le(n - 1\ri) \le[3n \le(n + 1\ri)\ri]^{n/(n-2)}} \,,
\ee
provided $n \neq 1, 2$ (see ref.
\cite{SDMFSS-PRD-2025}
for details).

The modified dynamics of the theory is shown to be derived from the corresponding equations of motion in an equivalent scalar-tensor formulation
\cite{SDMFSS-PRD-2025}.
In fact, while the violation of the EEP in the context of scalar-tensor theories has been studied by various authors in the past
\cite{ohanian-epviol,FTBM-epviol}, 
our formulation is arguably among the simplest and most concrete.

\section{Conclusions} \label{sec:concl}

Let us conclude with a discussion on the following:

\vspace{4pt}
\no 
Firstly, the singularity freeness of theory, which is mentioned above in section\,\ref{sec:cov}, right after Eq.\,(\ref{Phi}) that shows the finiteness of the function $\Phi (r)$, and hence of the modified gravitational force given by Eq.\,(\ref{grav2}), as $r \to 0$. While the singularity problem is a long-standing one
\cite{akr-PR-1955, penrose-PRL-1965, PH-ProcRS-1970},
%
our proposal here, which pertains to the formalism in ref.
\cite{SDMFSS-PRD-2025},
provides a succinct aversion of the same, albeit within the classical realm. 
%

\vspace{4pt}
\no 
Secondly, the true novelty of the proposal, as being that of a  EEP violation in a {\em metric theory} of gravity, which concerns {\it only} the gravitational couplings of matter fields. Although this may seem otherwise, from our allusion to a scalar-tensor equivalent setup for our formalism, it must be emphasized that such an equivalence pertains {\it only} to a $f(R)$ gravity scenario, which is nonetheless purely gravitational. So, there is no presumption of any additional (non-gravitational) scalar field as such, to which the matter fields can couple. This may further be reckoned from what we have stated right after Eq.\,(\ref{grav2}) in section\,\ref{sec:EEPviol} -- that our modified force law cannot conform to the standard one by a mere redefinition of $G$ or $\Phi(r)$, which implies that our proposal has no direct bearing on a varying $G$ theory or the MOdified Newtonian Dynamics (MOND) theory. In fact, a varying $G$ may still be used to account for Eq.\,(\ref{grav2}) under certain stipulations, for e.g., from its emergence from a {\em non-local} teleparallel equivalent GR of a specific sort
\cite{HM-NLG,BCHM-NLG,mash-NLG},
which we have studied in an earlier work concerning a unified cosmological picture without a stringent need for dark matter or(and) dark energy
\cite{SDSS-EPJP-2024}.
Nevertheless, the MOND-like approach to the dark matter is simply not that fundamental, as it introduces an acceleration scale rather than a length scale. 

\vspace{4pt}
\no 
Finally, the various length scales involved in our proposal -- the three distinct ones, viz., $\ell$, $\l$ and $\bar \l \sim 1/m$, as can be seen from Eqs.\,(\ref{mg3}) and (\ref{pot}), and the universality of the constant $m$. Among these three, $\ell$ represents the length scale at which deviations from the inverse square law of gravity occur. It must therefore be bounded as $\ell \lesssim 10^{-5}$\,m, since the inverse square law has been tested down to that scale
\cite{LACFH-submm}.
On the other hand, since the EEP violation may occur at practically any length scale, there are no intrinsic bounds on $\l$ and $\bar \l$. In principle, all the three scales may coincide, and potentially be equal to the Planck scale (which may be compared with that for the strong equivalence condition violation in certain contexts, see for e.g., ref.
\cite{SW-PRL-2011}). 
We intend to explore such a possibility, along with other plausible tests of our proposal, in a future work.

\bigskip
\section*{Acknowledgement}
This work is supported by the Natural Sciences and Engineering Research Council of Canada.
SS acknowledges financial support from the Faculty Research Programme Grant -- IoE, University of Delhi (Ref.No./ IoE/ 2025-26/ 12/ FRP).

\end{document}